\RequirePackage{ifpdf}

\RequirePackage{fixltx2e}
\RequirePackage{fix-cm}

\documentclass[%
floatfix,
showkeys,
nofootinbib, %
superscriptaddress, %
]{revtex4-1}

\usepackage{iftex}
\ifpdftex
\usepackage[noTeX]{mmap}	
\fi

\usepackage{textcomp}

\usepackage[T1,T2A]{fontenc}

\usepackage[utf8]{inputenx}
\input{ix-utf8enc.dfu}

\usepackage[german,russian,english]{babel}

\usepackage{amsmath}
\usepackage{amssymb}

\usepackage{mathtools}
\mathtoolsset{
showonlyrefs,
mathic = true
}

\allowdisplaybreaks

\usepackage{hyperref}
\hypersetup{backref,
 colorlinks=false}
\hypersetup{pdfborder=0 0 0}

\usepackage{microtype}
\UseMicrotypeSet[protrusion]{alltext}

\usepackage{graphicx}

\usepackage[scanall]{psfrag}

\usepackage{listings}
\usepackage{listingsutf8}
\usepackage[frozencache]{minted}

\setminted{breaklines=true}

\usepackage{fixltx2e}

\usepackage{nicefrac}

\makeatletter
\def\ps@pprintTitle{%
     \let\@oddhead\@empty
     \let\@evenhead\@empty
     \let\@oddfoot\@empty
     \let\@evenfoot\@oddfoot}
\makeatother

\usepackage{physics}

\begin{document}

\graphicspath{{image/asymptote-animation/en/}{image/asymptote-animation/}{image/}}

\title{Asymptote-based scientific animation}

\author{Migran N. Gevorkyan}
\email{gevorkyan-mn@rudn.ru}
\affiliation{RUDN University, \\
6 Miklukho-Maklaya St, Moscow, 117198, Russian Federation}

\author{Anna V. Korolkova}
\email{korolkova-av@rudn.ru}
\affiliation{RUDN University, \\
6 Miklukho-Maklaya St, Moscow, 117198, Russian Federation}

\author{Dmitry S. Kulyabov}
\email{kulyabov-ds@rudn.ru}
\affiliation{RUDN University, \\
6 Miklukho-Maklaya St, Moscow, 117198, Russian Federation}
\affiliation{Joint Institute for Nuclear Research,\\
  6 Joliot-Curie, Dubna, Moscow region, 141980, Russian Federation}

\begin{abstract}
  This article discusses a universal way to create animation using Asymptote the language for vector graphics. The Asymptote language itself has a built-in library for creating animations, but its practical use is complicated by an extremely brief description in the official documentation and unstable execution of existing examples. The purpose of this article is to eliminate this gap. The method we describe is based on creating a PDF file with frames using Asymptote, with further converting it into a set of PNG images and merging them into a video using FFmpeg. All stages are described in detail, which allows the reader to use the described method without being familiar with the used utilities.
\end{abstract}


\maketitle

\section{Introduction}
\label{sec:intro}

In this paper we study the creation of animation animation using the vector graphics language Asymptote~\cite{bowman_2012_surface-parameterization_en, bowman_2010_asymptote-interactive-graphics_en, bowman_2008_asymptote-vector-graphics-language_en, bowman_site_asymptote_en}.

Asymptote is an interpreted language, that is a translator into the PostScript vector graphics language. Designed to create vector images for mathematical publications. It is closely integrated with the \TeX{} system and is an integral part of the \TeX{} Live~\cite{texlive} distribution. It has a \verb|C|-like syntax, supports the creation of functions, custom data structures, and comes with an extensive set of modules for various tasks. Unlike PGF/TikZ~\cite{pgftikz}, Asymptote is more imperative, so it is easier to implement complex program logic on it.
  
In the official documentation of this language, only a few paragraphs are devoted to the animation creation process and the user is referred to the source code examples located in the \verb|animations| directory.

Asymptote creates animation in two steps. At the first step, a multi-page PDF file is created containing images that will become frames of future animation. Then, using the external utility Imagemagick~\cite{imagemagick} (command \verb|convert|), this PDF file is converted into a GIF image. If the Imagemagick utility is not installed on the user's system, all examples will stop at creating a multi-page pdf file with a set of images and a GIF image with animation will not be received.

In this article, we are considering a universal way to create animation in video format using the ffmpeg~\cite{ffmpeg:tomar2006,ffmpeg:site} and Ghostscript~\cite{ghostscript} utilities. All external programs will be called explicitly from the command line. With the help of Asymptote, only a multi-page PDF file with frames for the future video will be created.

The reader should be familiar with the basic capabilities of the Asymptote language. For an introduction to the basics of the language, we recommend the manual~\cite{Asymptote:tutorial}. The information from it will be enough to understand this work. 

\subsection{Article structure}
\label{sec:structure}

As an example, we chose the animation of the process of constructing epitrochoids and hypotrochoids. In the first part of the work, we will recall the definitions of these curves, some of their properties and reduce their construction to a composition of two rotations. In the second part of the article, we will describe in detail the implementation of their construction using Asymptote. And in the third part we will focus on the technical side of the issue and describe the process of creating a multi-page PDF file, converting it into PNG images using Ghostscript and converting these images into video using ffmpeg.

  \section{Task description}

  Consider the task of animating the process of constructing cycloidal curves, namely hypotrochoids and epitrochoids. We will not use the parametric equation of these curves, but will reduce everything to the composition of two rotations applied to the starting point of the curve. This will better illustrate the capabilities of the Asymptote language.

  \subsection{Definition of epitrochoids}

\begin{figure}[h!]
  \begin{minipage}[b]{0.45\textwidth}
    \centering
    \includegraphics{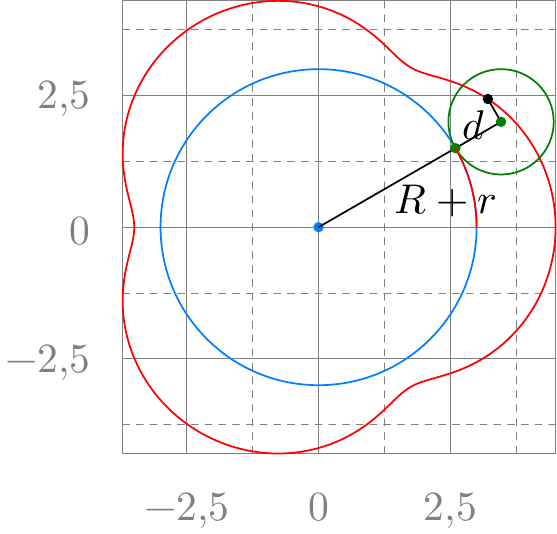}
    \caption{$R=3, r=1, d=1/2$}
    \label{fig:01}
  \end{minipage}
  \begin{minipage}[b]{0.45\textwidth}
    \centering
    \includegraphics{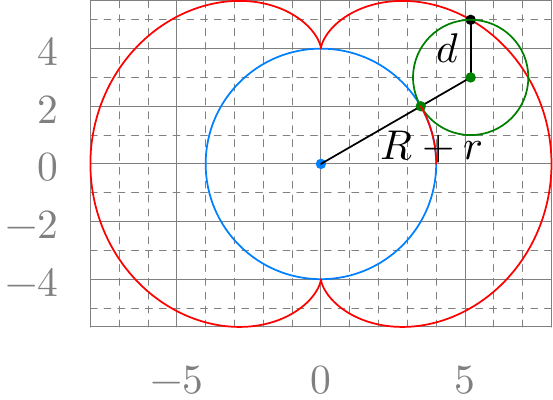}
    \caption{$R=4, r=2, d=2$}
    \label{fig:02}
  \end{minipage}
\end{figure}

  \emph{Epitrochoid} is defined as a trajectory plotted by a fixed point $P$ lying on a radial line of circle with radius $r$, which rolls along the \emph{outer side} of the circle with radius $R$ (fig.~\ref{fig:01}). The parametric equation of the curve has the following form:
\begin{equation*}
  \left\{
    \begin{aligned}
      &x(t) = (R+r)\cos(\varphi) - d\cos\Big(\frac{R+r}{r}\varphi\Big),\\
      &y(t) = (R+r)\sin(\varphi) - d\sin\Big(\frac{R+r}{r}\varphi\Big),\\
    \end{aligned}
  \right.
\end{equation*}
where $d$ is the distance from the center of the rolling circle to the point of the curve, $\varphi$ is the angle of rotation of the rolling circle relative to the axis $Ox$.

Let use introduce the coefficient $k = r/R$, then it will possible to change the parameterization and the equation will take the form:
\begin{equation*}
  \left\{
    \begin{aligned}
      &x(t) = R(k+1)\cos(kt) - d\cos((k+1)t),\\
      &y(t) = R(k+1)\sin(kt) - d\sin((k+1)t),\\
    \end{aligned}
  \right.
\end{equation*}
where the parameters $t$ and $\varphi$ are related as $\varphi = kt$.

Some special cases of epitrochoids have proper names. So for $r = R$, \emph{Pascal's snail} is obtained, for $d = R+r$ ---\emph{rosy curve} or \emph{rose}, and for $d=r$ --- \emph{epicycloid} (fig.~\ref{fig:02}).

\subsection{Definition of a hypotrochoid}

\begin{figure}[h!]
  \begin{minipage}[b]{0.45\textwidth}
    \centering
    \includegraphics{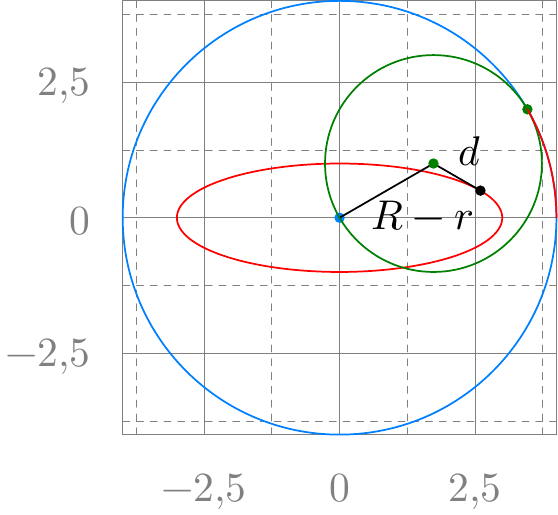}
    \caption{$R=4, r=2, d=1$}
    \label{fig:03}
  \end{minipage}
  \begin{minipage}[b]{0.45\textwidth}
    \centering
    \includegraphics{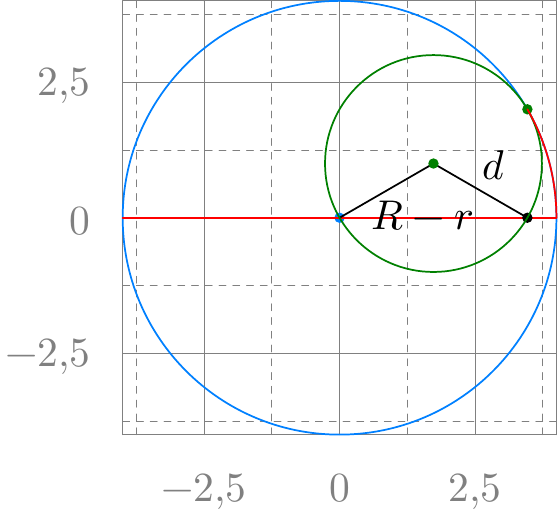}
    \caption{$R=4, r=2, d=2$}
    \label{fig:04}
  \end{minipage}
\end{figure}

  \emph{Hypotrochoid} --- the trajectory described by a fixed point $P$ on a radial straight circle of radius $r$, which rolls along the \emph{inner} side of the circle of radius $R$ (fig.~\ref{fig:03}). The parametric equation of the curve has the following form:
\begin{equation*}
  \left\{
    \begin{aligned}
      &x(t) = (R-r)\cos(\varphi) + d\cos\Big(\frac{R-r}{r}\varphi\Big),\\
      &y(t) = (R-r)\sin(\varphi) - d\sin\Big(\frac{R-r}{r}\varphi\Big),\\
    \end{aligned}
  \right.
\end{equation*}
  where, as in the case of the epitrochoid, $d$ is the distance from the center of the rolling circle to the point $P$. In particular, for $d=r$, \emph{hypocycloid} is obtained (fig.~\ref{fig:04}).

  It is also possible to parametrize $\varphi = kt$, where $k = r/R$, then the equation will take the form:
\begin{equation*}
  \left\{
    \begin{aligned}
      &x(t) = R(1-k)\cos(kt) + d\cos((1-k)t),\\
      &y(t) = R(1-k)\sin(kt) - d\sin((1-k)t),\\
    \end{aligned}
  \right.
\end{equation*}

  \subsection{Reducing the problem to a composition of turns}

  The construction of cycloidal curves begins by specifying two circles: a fixed circle of radius $R$ centered at point $O_R$ and a moving circle of radius $r$ centered at point $O_r$.

  A fixed circle will be conventionally called "large", and a moving one --- "small", since usually $R > r$. On the radial line of a small circle, we fix the point of the curve $P_0$.
  
\begin{figure}[h!]
    \begin{minipage}[b]{0.32\textwidth}
      \centering
      \includegraphics{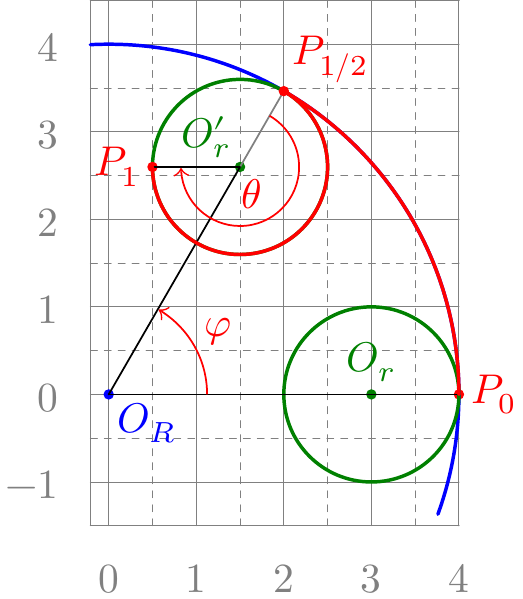}
      \caption{Hypocycloid $d = r$}
      \label{fig:05}
    \end{minipage}
    \begin{minipage}[b]{0.32\textwidth}
      \centering
      \includegraphics{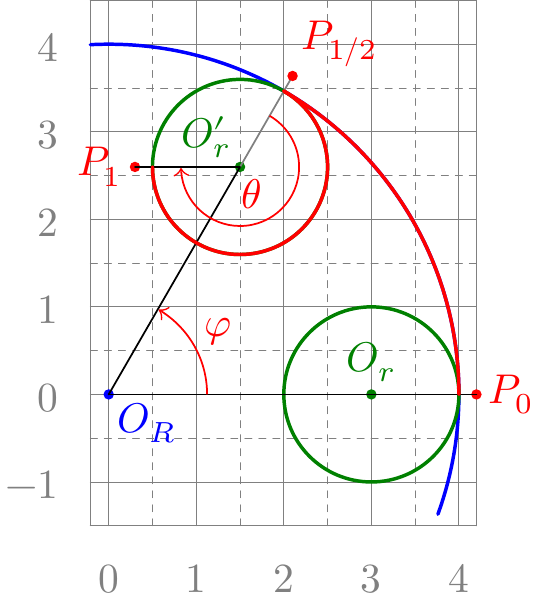}
      \caption{Hypotrochoid $d > r$}
      \label{fig:06}
    \end{minipage}
    \begin{minipage}[b]{0.32\textwidth}
      \centering
      \includegraphics{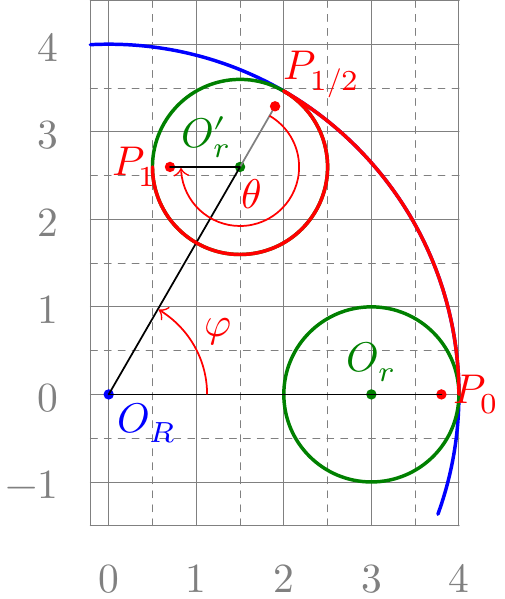}
      \caption{Hypotrochoid $d < r$}
      \label{fig:07}
    \end{minipage}
  \end{figure}

  From the definition of hypotrochoids and epitrochoids, it follows that a motion $T(\varphi)$ is performed over the point $P_0$, consisting of a composition of two turns~(fig.~\ref{fig:05}--\ref{fig:07}).
  \begin{enumerate}
    \item $T_1(\varphi)$ --- rotation around the point $O_R$ by the angle $\varphi$, at which the point $O_r$ turns into $O^{\prime}_{r}$, and the point $P_0$ into the point $P_{1/2}$.
    \item $T_2(\theta(\varphi))$ --- rotation around the point $O^{\prime}_{r}$ by the angle $\theta$, at which $P_{1/2}$ turns into $P_{1}$.
  \end{enumerate}

  \begin{figure}[h!]
    \begin{minipage}[b]{0.32\textwidth}
      \centering
      \includegraphics{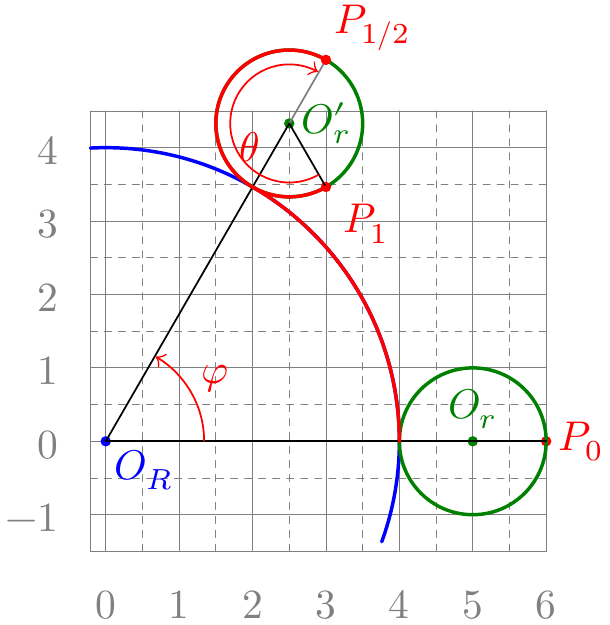}
      \caption{Epicycloid $d = r$}
      \label{fig:08}
    \end{minipage}
    \begin{minipage}[b]{0.32\textwidth}
      \centering
      \includegraphics{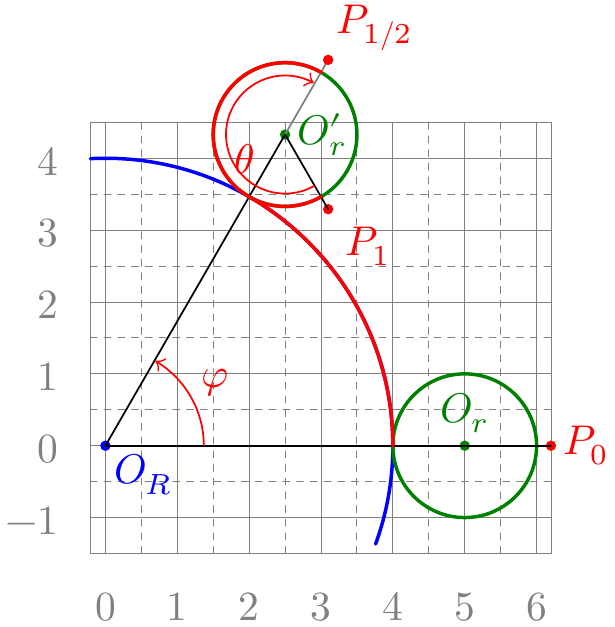}
      \caption{Epitrochoid $d > r$}
      \label{fig:09}
    \end{minipage}
    \begin{minipage}[b]{0.32\textwidth}
      \centering
      \includegraphics{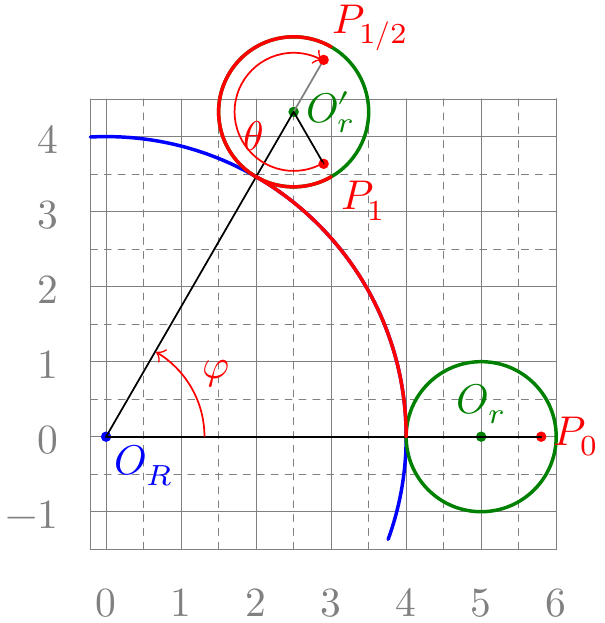}
      \caption{Epitrochoid $d < r$}
      \label{fig:10}
    \end{minipage}
  \end{figure}

  The rotation angle $\theta$ is related to the angle $\varphi$. A small circle must travel a distance equal to the length of the arc $PP_{1/2}$, which means that the lengths of the arcs $PP_{1/2}$ and $P_{1/2}P_{1}$ are equal.
\begin{equation*}
  |PP_{1/2}| = R\varphi = |P_{1/2}P_{1}| = \theta r \Rightarrow \theta = \dfrac{R\varphi}{r} = \dfrac{\varphi}{k},\; k = r/R.
\end{equation*}

  Thus, to construct a curve, it is enough to set the parameters $R$, $r$ and $d$, the initial positions of the circles and the points $P_0$. It is usually assumed that the center of $O_R$ coincides with the origin, and the center of $O_r$ lies on the $Ox$ axis. Then the coordinates of the center $O_r$ are calculated as:
  \begin{equation}\label{eq:01}
    \vb{OO}_r = \vb{OO}_R + (R + s \cdot r, 0)^T,\quad
    s = 
    \left\{
      \begin{aligned}
        +1 & \text{, if epitrochoid},\\
        -1 & \text{, if hypotrochoid}.\\
      \end{aligned}
    \right.
  \end{equation}
  And the coordinates of the curve point are $P_0$:
\begin{equation*}
  \vb{OP}_{0} = \vb{OO}_r + (d, 0)^T.
\end{equation*}

  Now, to find any point of the curve, it is enough to act on $P_0$ by converting $T(\varphi) = T_2(\varphi/k) \circ T_1(\varphi)$ setting the required value to $\varphi$. If it is necessary to construct a set of points, then taking a sufficiently small step $\varphi$, one can consistently act on the point $P_0$ by transformations $T(i\varphi),i=1,2,\ldots,n$:
\begin{equation*}
  P_0 \stackrel{T( \varphi)}{\longrightarrow} P_1,\;
  P_0 \stackrel{T(2\varphi)}{\longrightarrow} P_2,\;
  P_0 \stackrel{T(3\varphi)}{\longrightarrow} P_3,\;
  P_0 \stackrel{T(4\varphi)}{\longrightarrow} P_4,\;,\ldots,
  P_0 \stackrel{T(n\varphi)}{\longrightarrow} P_n,\;.
\end{equation*}

  \section{Implementation on Asymptote}

Below we present the source code of the program in the Asymptote language and comment on its key points.

\newcounter{codlabel}
\newcommand{\cod}[1]{\refstepcounter{codlabel}(\thecodlabel)\label{#1}}

\begin{minted}{asymptote}
include "config.asy";

import animation;
import graph;

unitsize(1cm);
size(10cm, 10cm);

string ssign(int d) {
  return d > 0 ? "+" : "-";
}

transform T1(real phi, pair O_R) { // \cod{m1}
  return rotate(angle=phi, z=O_R);
}

transform T2(real phi, int sign, real k, pair O_r) { // \cod{m2}
  return rotate(angle=sign*phi/k, z=O_r);
}

pen BigCircle = blue;
pen littleCircle = deepgreen;
pen curve = red;

int sign = -1; // \cod{m3}
real R = 4.0;
real r = 2.0;
real d = 2.0;
real k = r/R;
int N = 100; // \cod{m4}
pair O_R = (0, 0);
int turns = 1; // \cod{m5}
usersetting(); // \cod{m6}

pair O_r = O_R + polar(R + sign * r, 0); // \cod{m7}
pair P = O_r + polar(d, 0);  // \cod{m8}
pair Q = O_r - sign * polar(r, 0);  // \cod{m9}

guide xcycloid;
transform T;

animation A; // \cod{m10}
A.global = true; 

draw(circle(c=O_R, r=R), p=BigCircle); // \cod{m11}
dot(O_R, p=BigCircle);


for(real phi: uniform(0, 360turns, N)) {
  save(); // \cod{m12}
  
  T = T1(phi, O_R) * T2(phi, sign, k, O_r); // \cod{m13}
  
  xcycloid = xcycloid -- T*P; // \cod{m14}

  draw(xcycloid, p=1bp+curve); // \cod{m15}
  dot(T*P, L=Label("P", align=NW)); // \cod{m16}
  draw(O_R -- T*O_r, L=Label("R"+ssign(sign)+"r")); // \cod{m17}
  draw(T*O_r -- T*P, L=Label("d"));
  draw(circle(c=T*O_r, r=r), p=littleCircle); // \cod{m18}
  dot(T*O_r, p=littleCircle);
  dot(T1(phi, O_R)*Q, p=littleCircle); // \cod{m19}

  include "axes.asy"; // \cod{m20}

  A.add(); // \cod{m21}
  restore(); // \cod{m22}
}

A.movie(); // \cod{m23}
currentpicture.erase();

\end{minted}

This program creates a multi-page PDF file, each page of which is a future frame of the video. The main work on calculating the points of the curve is performed by the functions \verb|T1|~\eqref{m1} and \verb|T2|~\eqref{m2}. These functions are defined for convenience, so that the code reflects the above formulas as much as possible. All the work is done by the built-in function \verb|rotate|, which allows you to determine rotation around an arbitrary point (argument \verb|z|) by an arbitrary angle value in degrees (argument \verb|angle|).

Next, we set a set of variables-parameters ~\eqref{m3}. The variable \verb|sign| --- is $s$ from the formula~\eqref{eq:01}, and the rest correspond to their mathematical notation. The variable \verb|N|~\eqref{m4} sets the number of calculated points and, as a result, frames in the future video. The variable \verb|turns|~\eqref{m5} sets the number of complete turns around the center of $O_R$. Calling the built-in function \verb|usersetting|~\eqref{m6} to override the value of any variable specified above via the command line argument \verb|-u|.

Then, based on the above-defined parameters, the coordinates of the initial position of the center of the moving circle $O_r$~\eqref{m7}, the points of the curve $P$~\eqref{m8} and the point of tangency $Q$ of the moving circle with the stationary ~\eqref{m9} are calculated.

Next, an object is created \verb|A|~\eqref{m10}, into which animation frames will be recorded (objects of the type \verb|picture| or \verb|frame|). \verb|A| has several fields, in particular the \verb|global| field of the \verb|bool| type allows you to enable and disable saving the created images as an array in RAM and writing them as files to disk only after they are all built.
  
The curve points are calculated in a loop, but before that, a fixed circle~\eqref{m11} and its center are drawn. Then, at the beginning of each iteration of the cycle, all the current stationary elements of the image are saved (object \verb|picture|)~\eqref{m12}, all movable elements are built, the resulting image is added to the structure \verb|A|~\eqref{m21} and the image state is reset~\eqref{m22} to the which was at the time of ~\eqref{m12}. The process continues until all frames are drawn and saved to \verb|A|.
  
As the cycle progresses, the angle $\varphi$ changes from $0$ to $2\pi n$ (in degrees). At each step, the rotation transformation $T(\varphi)$ is calculated~\eqref{m14}, applied to the starting point of $P$ and added to the path (\verb|guide|) \verb|xcycloid|~\eqref{m14}. With each iteration of the loop, new points are added to the path \verb|xcycloid| and the curve grows.

The following drawing commands follow:
\begin{itemize}
\item of the already calculated part of the curve~\eqref{m15};
\item of the new point position $P$~\eqref{m16};
\item of a segment of length $R +s\cdot r$~\eqref{m17} connecting the center of $O_R$ to the new position of the center of $O_r$, as well as a segment of length $d$ connecting the new center of $O_r$ to the point $P$ of the curve;
\item directly the moving circle itself in its new position~\eqref{m18} and its center;
\item touch point $Q$~\eqref{m19};
\item coordinate grid, the settings of which are placed in a separate file~\eqref{m20}.
\end{itemize}

Finally, after working out the loop, all created frames are recorded in a PDF file. To do this, Asymptote sequentially creates separate PDF files for each frame, then adds text processed by \LaTeX{} (in our case Lua\LaTeX{}) to them. It is this procedure that takes the main time of the program, the calculations themselves practically do not take up time in comparison with this.

We also note the peculiarity of the Asymptote syntax, which allows omitting the \verb|*| operator when multiplying numeric literal constants and variables, for example \verb|360turn|~\eqref{m12}.
  \section{Creating a video clip}
  \subsection{Launching Asymptote}

  To run the program discussed above, run the following command.
\begin{minted}[breaklines=true,linenos=false]{bash}
  asy -noV -nobatchView -f pdf -globalwrite -u 'R=3;r=1;d=1;N=100' xcycloid.asy -o video/xcycloid.pdf
\end{minted}

  The source code file \verb|xcycloid.asy| is started for execution and as a result the file \verb|xcycloid.pdf| will be created. Consider the options used.
  \begin{itemize}
    \item Options \verb|-noV| and \verb|-nobatchView| prevent the newly created image from opening automatically. The \verb|-noV| option disables this function when executed from the command line, and \verb|-nobatchView| when executing the script (as in our case).
    \item Option \verb|-f pdf| indicates that you should immediately create a PDF file, bypassing the postscript file stage.
    \item option \verb|-globalwrite| makes it possible to save the file \verb|xcycloid.pdf| to any directory (in our case \verb|video|), and not only to the one where the source file \verb|xcycloid.asy| is located.
    \item Option \verb|-u| allows you to interact with the \verb|usersetting()| function and pass the values of variables inside the program. So we pass the values \verb|R=3|, \verb|r=1|, \verb|d=1| and \verb|N=100|. This feature allows you to use a single source code file to build multiple images, flexibly adjusting any parameters. Note that this parameter takes exactly a text string, which is then processed by the \verb|usersetting()| function, so the passed parameters must be taken in quotation marks.
    \end{itemize}

  \subsection{Converting to PNG using GhostScript}

  To convert the resulting multipage file into a video format, it is necessary to convert its pages into bitmaps. To do this, we suggest using the GhostScript~\cite{ghostscript} program. It is available for both Windows and Unix systems (GNU/Linux, macOS). It also comes with the \TeX{}Live~\cite{texlive} distribution, as does Asymptote/

  To convert a PDF file, run the command
\begin{minted}[breaklines=true,linenos=false]{bash}
  gs -sDEVICE=png16m -r600 -o video/xcycloid-%04d.png video/xcycloid.pdf
\end{minted}

  In the case of using GhostScript from the \TeX{}Live distribution, you should call \verb|gs| using the \verb|rungs| script, which is located
  \begin{itemize}
    \item in the directory \verb|texlive\2023\bin\win32| in the case of Windows OS,
    \item in the \verb|texlive/2023/bin/x86_64-linux| in the case of GNU/Linux.
  \end{itemize}
  The \verb|2023| directory corresponds to the version of the \TeX{}Live distribution and may differ.

  \subsection{Creating a video using FFmpeg}

  The process of gluing the resulting bitmap images into one video clip is carried out using FFmpeg~\cite{ffmpeg:site}. This program is a command-line utility and has extensive functionality and, as a result, a huge number of options and settings. Let's give an example of creating a video clip from the PNG images generated in the previous step and give an explanation of the parameters used.
\begin{minted}[breaklines=true,linenos=false]{bash}
  ffmpeg -r 30 -f image2 -start_number 1 -i video/xcycloid-%04d.png -c:v libx264 -vf "pad=ceil(iw/2)*2:ceil(ih/2)*2" video/xcycloid.mp4
\end{minted}

  \begin{itemize}
    \item parameter \verb|-r| sets the frame rate.
    \item parameter \verb|-f| sets the format of the input file.
    \item Since a lot of files are submitted to the input, you should specify the format of their names. The same notation is used as in the case of \verb|gs|. The \verb|-start_number| parameter sets the starting number.
    \item Parameter \verb|-c:v| allows you to specify the video encoder used. In our case \verb|libx264|, but many other formats are supported.
    \item The important parameter \verb|-vf| sets the filter that is applied to the processed frame. In our case, we round the width and height of the frame in pixels to an even number. After converting to PNG, the width and height of the image may be odd, which is unacceptable for the vast majority of encoders. The specified filter allows you to avoid this error and rescale the frame by ffmpeg.
  \end{itemize}
  At the output we will get a video packed in a container \verb|mp4|. The x264 format we have chosen is widespread and can be played by any modern browser, not to mention video player programs.
\section{Conclusion}
\label{sec:conclusion}

  We have analyzed in detail the way to create vector graphics animation on a plane using the Asymptote language. This aspect of this language is poorly covered in the official manual and, in our opinion, this article fills this gap. Although the result is a video clip containing bitmaps, but thanks to the vector source (PDF), you can increase the resolution of the video almost limitlessly.

  It should also be noted that this method of creating animation is universal, since almost any data visualization tool can be used to create a set of image frames. FFmpeg does all the work on creating a video file.

\begin{acknowledgments}

This paper has been supported by the RUDN University Strategic Academic Leadership Program.

\end{acknowledgments}

 \bibliographystyle{elsarticle-num-names}

\bibliography{bib/asymptote-animation/cite}

\begin{thebibliography}{10}
\def\selectlanguageifdefined#1{
\expandafter\ifx\csname date#1\endcsname\relax
\else\selectlanguage{#1}\fi}
\providecommand*{\href}[2]{{\small #2}}
\providecommand*{\url}[1]{{\small #1}}
\providecommand*{\BibUrl}[1]{\url{#1}}
\providecommand{\BibAnnote}[1]{}
\providecommand*{\BibEmph}[1]{#1}
\ProvideTextCommandDefault{\cyrdash}{\iflanguage{russian}{\hbox
  to.8em{--\hss--}}{\textemdash}}
\providecommand*{\BibDash}{\ifdim\lastskip>0pt\unskip\nobreak\hskip.2em plus
  0.1em\fi
\cyrdash\hskip.2em plus 0.1em\ignorespaces}
\renewcommand{\newblock}{\ignorespaces}

\bibitem{bowman_2012_surface-parameterization_en}
\selectlanguageifdefined{english}
\BibEmph{Shardt~O., Bowman~J.~C.} Surface parameterization of nonsimply
  connected planar Bézier regions~//
  \href{http://dx.doi.org/10.1016/j.cad.2011.05.010}{\BibEmph{Computer-Aided
  Design}}. \BibDash
\newblock 2012. \BibDash May. \BibDash
\newblock Vol.~44, no.~5. \BibDash
\newblock P.~484.e1--484.e10.

\bibitem{bowman_2010_asymptote-interactive-graphics_en}
\selectlanguageifdefined{english}
\BibEmph{Bowman~J.~C.} Asymptote: Interactive TEX-aware 3D vector graphics.
  \BibDash
\newblock 2010. \BibDash
\newblock Vol.~31, no.~2. \BibDash
\newblock P.~203--205.

\bibitem{bowman_2008_asymptote-vector-graphics-language_en}
\selectlanguageifdefined{english}
\BibEmph{Bowman~J.~C., Hammerlindl~A.} Asymptote: A vector graphics language.
  \BibDash
\newblock 2008. \BibDash
\newblock Vol.~29, no.~2. \BibDash
\newblock P.~288--294.

\bibitem{bowman_site_asymptote_en}
\selectlanguageifdefined{english}
\BibEmph{Bowman~J.~C.} Asymptote: The Vector Graphics Language. \BibDash
\newblock 2023. \BibDash 5. \BibDash
\newblock Access mode: \BibUrl{https://asymptote.sourceforge.io/}.

\bibitem{texlive}
\selectlanguageifdefined{english}
{TeX Live}. \BibDash
\newblock 2023. \BibDash
\newblock Access mode: \BibUrl{https://www.tug.org/texlive/}.

\bibitem{pgftikz}
\selectlanguageifdefined{english}
\BibEmph{Tantau~T., Menke~H.} {PGF/TikZ}. \BibDash
\newblock 2023. \BibDash
\newblock Access mode: \BibUrl{https://ctan.org/pkg/pgf}.

\bibitem{imagemagick}
\selectlanguageifdefined{english}
\BibEmph{{The ImageMagick Development Team}}. {ImageMagick}. \BibDash
\newblock 2023. \BibDash
\newblock Access mode: \BibUrl{https://imagemagick.org}.

\bibitem{ffmpeg:tomar2006}
\selectlanguageifdefined{english}
\BibEmph{Tomar~S.} {Converting video formats with FFmpeg}~// \BibEmph{Linux
  Journal}. \BibDash
\newblock 2006. \BibDash
\newblock Vol. 2006, no. 146. \BibDash
\newblock P.~10.

\bibitem{ffmpeg:site}
\selectlanguageifdefined{english}
{FFmpeg Website}. \BibDash
\newblock 2023. \BibDash
\newblock Access mode: \BibUrl{https://ffmpeg.org/}.

\bibitem{ghostscript}
\selectlanguageifdefined{english}
{Ghostscript Website}. \BibDash
\newblock 2023. \BibDash
\newblock Access mode: \BibUrl{https://www.ghostscript.com/}.

\bibitem{Asymptote:tutorial}
\selectlanguageifdefined{english}
\BibEmph{Staats~C.~I.} {An Asymptote tutorial}. \BibDash
\newblock 2015. \BibDash
\newblock Access mode:
  \BibUrl{https://math.uchicago.edu/~cstaats/Charles_Staats_III/Notes_and_papers_files/asymptote_tutorial.pdf}.

\bibitem{Asymptote:Kryachkov:2015}
\selectlanguageifdefined{russian}
\BibEmph{Крячков~Ю.~Г.} {Asymptote для начинающих}. \BibDash
\newblock 2015. \BibDash 07. \BibDash
\newblock {Режим доступа}: \BibUrl{http://mif.vspu.ru/books/ASYfb.pdf}.

\bibitem{Asymptote:Volchenko:2018}
\selectlanguageifdefined{russian}
\BibEmph{Волченко~Ю.~М.} {Научная графика на языке Asymptote}. \BibDash
\newblock 2018. \BibDash 02. \BibDash
\newblock {Режим доступа}: \BibUrl{http://www.math.volchenko.com/AsyMan.pdf}.

\end{thebibliography}

\end{document}


\graphicspath{{image/asymptote-animation/ru/}{image/asymptote-animation/}{image/}}

\title{Научная анимация на основе Asymptote}

\author{М. Н. Геворкян}
\email{gevorkyan-mn@rudn.ru}
\affiliation{Российский университет дружбы народов, \\
117198, Москва, ул. Миклухо-Маклая, д. 6}

\author{А. В. Королькова}
\email{korolkova-av@rudn.ru}
\affiliation{Российский университет дружбы народов, \\
117198, Москва, ул. Миклухо-Маклая, д. 6}

\author{Д. С. Кулябов}
\email{kulyabov-ds@rudn.ru}
\affiliation{Российский университет дружбы народов, \\
117198, Москва, ул. Миклухо-Маклая, д. 6}
\affiliation{Объединённый институт ядерных исследований,\\
  ул. Жолио-Кюри 6, Дубна, Московская область, Россия, 141980}

\begin{abstract}

  В данной статье рассматривается универсальный способ создания анимации с помощью языка для создания векторной графики Asymptote. В сам язык Asymptote встроена библиотека для создания анимации, однако практическое ее использование осложнено крайне кратким описанием в официальной документации и нестабильной работой существующих примеров. Целью данной статьи является устранение данного пробела. Излагаемый нами способ основывается на создании PDF файла с кадрами с помощью Asymptote, с дальнейшей конвертацией его в набор PNG изображений и склейкой их в видео с помощью FFmpeg. Все этапы подробно описываются, что дает возможность читателю использовать изложенный метод, не будучи знакомым с используемыми утилитами.

\end{abstract}

  \keywords{векторная графика,
    TeX,
    asymptote,
    научная графика}

\maketitle

\section{Введение}
\label{sec:intro}

В данной работе рассматривается способ создания анимации с помощью языка векторной графики Asymptote~\cite{bowman_2012_surface-parameterization_en, bowman_2010_asymptote-interactive-graphics_en, bowman_2008_asymptote-vector-graphics-language_en, bowman_site_asymptote_en}.

Asymptote --- интерпретируемый язык, представляющий собой транслятор в язык векторной графики PostScript. Предназначен для создания векторных изображений математической направленности. Тесно интегрирован с системой \TeX{} и является составной частью дистрибутива \TeX{} Live~\cite{texlive}. Имеет C-подобный синтаксис, поддерживает создание функций, пользовательских структур данных и поставляется с обширным набором модулей для разных задач. В отличии от PGF/TikZ~\cite{pgftikz} Asymptote более императивный, поэтому на нем проще реализовать сложную программную логику.

В официальной документации данного языка процессу создания анимации посвящено всего несколько абзацев и пользователя отсылают к примерам исходного кода, находящимся в директории \verb|animations|.

Asymptote создает анимацию в два этапа. На первом этапе создается многостраничный PDF-файл, содержащий изображения, которые станут кадрами будущей анимации. Затем с помощью внешней утилиты Imagemagick~\cite{imagemagick} (команда \verb|convert|), данный PDF-файл преобразуется в GIF-изображение. В случае, если утилита Imagemagick не установлена в системе пользователя, все примеры будут останавливаться на создании многостраничного pdf-файла с набором картинок и GIF-изображение с анимацией получено не будет.

В данной статье мы рассматриваем универсальный способ создания анимации в видео-формате с использованием утилит ffmpeg~\cite{ffmpeg:tomar2006,ffmpeg:site} и Ghostscript~\cite{ghostscript}. Все внешние программы будут вызываться явным способом из командной строки. С помощью Asymptote будет создаваться только многостраничный PDF-файл с кадрами для будущего видео.

От читателя требуется знакомство с базовыми возможностями языка Asymptote.  Для введения в основы языка рекомендуем пособие~\cite{Asymptote:tutorial}. Сведений из него будет достаточно для понимания данной работы. На русском языке также существует несколько пособий:~\cite{Asymptote:Kryachkov:2015} и~\cite{Asymptote:Volchenko:2018}.

\subsection{Структура статьи}
\label{sec:structure}

В качестве примера мы выбрали анимирование процесса построения эпитрохоид и гипотрохоид. В первой части работы мы напомним определения этих кривых, некоторые их свойства и сведем их построение к композиции из двух поворотов. Во второй части статьи подробно опишем реализацию их построения с помощью Asymptote. А в третей части остановимся на технической стороне вопроса и опишем процесс создания многостраничного PDF-файла, его конвертации в PNG-изображения с помощью Ghostscript и преобразование этих изображений в видео с помощью ffmpeg.

  \section{Описание задачи}

Рассмотрим задачу анимации процесса построения циклоидальных кривых, а именно гипотрохоиды и эпитрохоиды. Мы не будем пользоваться параметрическим уравнением данных кривых, а сведем все к композиции двух поворотов, применяемых к начальной точке кривой. Это позволит лучше проиллюстрировать возможности языка Asymptote.

  \subsection{Определение эпитрохоиды}

\begin{figure}[h!]
  \begin{minipage}[b]{0.45\textwidth}
    \centering
    \includegraphics{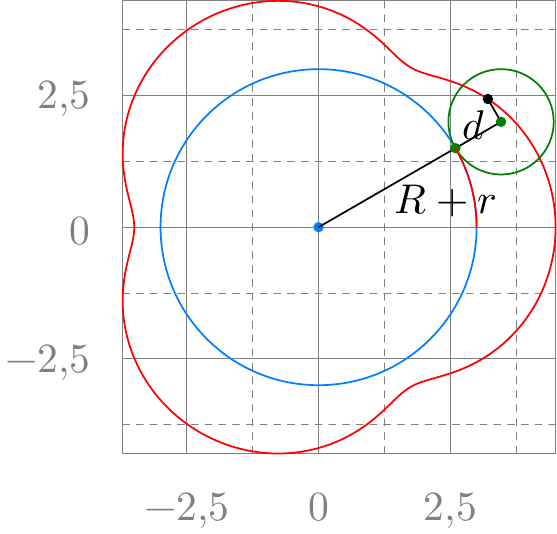}
    \caption{$R=3, r=1, d=1/2$}
    \label{fig:01}
  \end{minipage}
  \begin{minipage}[b]{0.45\textwidth}
    \centering
    \includegraphics{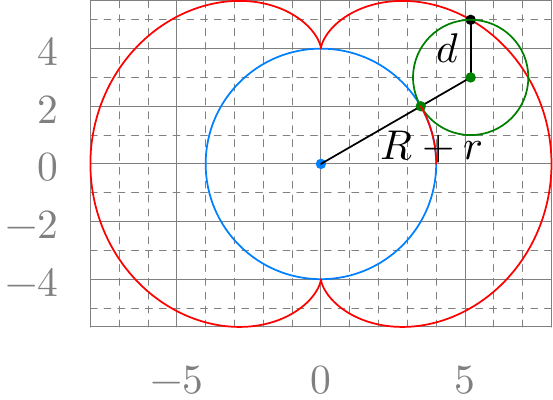}
    \caption{$R=4, r=2, d=2$}
    \label{fig:02}
  \end{minipage}
\end{figure}

\emph{Эпитрохоида} определяется как траектория, описываемся фиксированной точкой $P$, лежащей на радиальной прямой окружности радиуса $r$, которая катится по \emph{внешней стороне} окружности радиуса $R$ (рис.~\ref{fig:01}). Параметрическое уравнение кривой имеет следующий вид:
\begin{equation*}
  \left\{
    \begin{aligned}
      &x(t) = (R+r)\cos(\varphi) - d\cos\Big(\frac{R+r}{r}\varphi\Big),\\
      &y(t) = (R+r)\sin(\varphi) - d\sin\Big(\frac{R+r}{r}\varphi\Big),\\
    \end{aligned}
  \right.
\end{equation*}
где $d$ --- расстояние от центра катящейся окружности до точки кривой, $\varphi$ --- угол поворота катящейся окружности относительно оси $Ox$.

Если ввести коэффициент $k = r/R$, то можно поменять параметризацию и уравнение примет вид:  
\begin{equation*}
  \left\{
    \begin{aligned}
      &x(t) = R(k+1)\cos(kt) - d\cos((k+1)t),\\
      &y(t) = R(k+1)\sin(kt) - d\sin((k+1)t),\\
    \end{aligned}
  \right.
\end{equation*}
где параметры $t$ и $\varphi$ связаны как $\varphi = kt$.

Некоторые частные случаи эпитрохоиды имеют имена собственные. Так при $r = R$ получается \emph{улитка Паскаля}, при $d = R+r$ --- \emph{розовидная кривая} или \emph{роза}, а при $d=r$ --- \emph{эпициклоиду} (рис.~\ref{fig:02}).

\subsection{Определение гипотрохоиды}

\begin{figure}[h!]
  \begin{minipage}[b]{0.45\textwidth}
    \centering
    \includegraphics{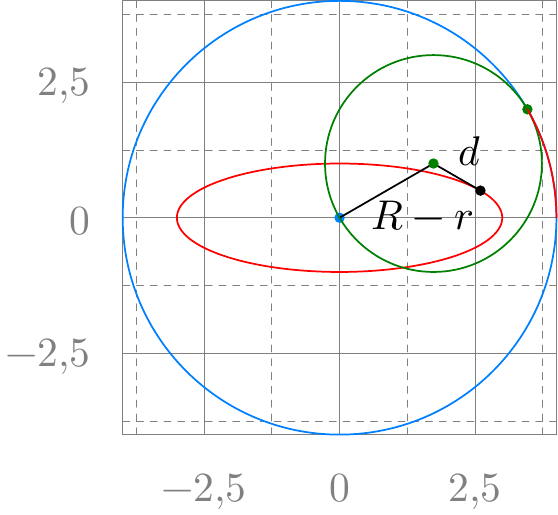}
    \caption{$R=4, r=2, d=1$}
    \label{fig:03}
  \end{minipage}
  \begin{minipage}[b]{0.45\textwidth}
    \centering
    \includegraphics{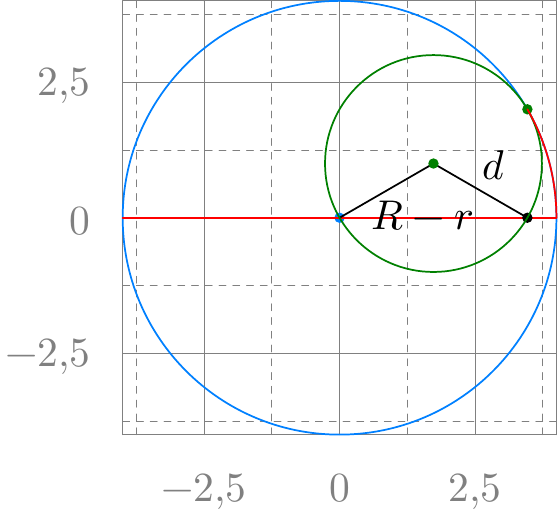}
    \caption{$R=4, r=2, d=2$}
    \label{fig:04}
  \end{minipage}
\end{figure}

  \emph{Гипотрохоида} --- траектория, которую описывает фиксированная точка $P$ на радиальной прямой окружности радиуса $r$, котора катится по \emph{внутренней} стороне окружности радиуса $R$ (рис.~\ref{fig:03}). Параметрическое уравнение кривой имеет следующий вид:
\begin{equation*}
  \left\{
    \begin{aligned}
      &x(t) = (R-r)\cos(\varphi) + d\cos\Big(\frac{R-r}{r}\varphi\Big),\\
      &y(t) = (R-r)\sin(\varphi) - d\sin\Big(\frac{R-r}{r}\varphi\Big),\\
    \end{aligned}
  \right.
\end{equation*}
  где, как и в случае эпитрохоиды, $d$ --- расстояние от центра катящейся окружности до точки $P$. В частности, при $d=r$ получается \emph{гипоциклоида} (рис.~\ref{fig:04}).

  Также возможна параметризация $\varphi = kt$, где $k = r/R$, тогда уравнение примет вид:
\begin{equation*}
  \left\{
    \begin{aligned}
      &x(t) = R(1-k)\cos(kt) + d\cos((1-k)t),\\
      &y(t) = R(1-k)\sin(kt) - d\sin((1-k)t),\\
    \end{aligned}
  \right.
\end{equation*}

  \subsection{Сведение задачи к композиции поворотов}

  Построение циклоидальных кривых начинается с задания двух окружностей: неподвижной окружности радиуса $R$ с центром в точке $O_R$ и движущейся окружности радиуса $r$ с центром в точке $O_r$.

  Неподвижную окружность будем условно называть «большой», а подвижную --- «малой», так как обычно $R > r$. На радиальной прямой малой окружности фиксируем точку кривой $P_0$.
  
\begin{figure}[h!]
    \begin{minipage}[b]{0.32\textwidth}
      \centering
      \includegraphics{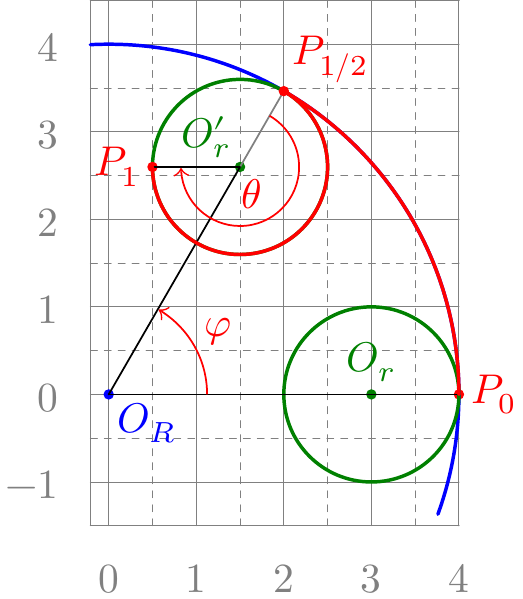}
      \caption{Гипоциклоида $d = r$}
      \label{fig:05}
    \end{minipage}
    \begin{minipage}[b]{0.32\textwidth}
      \centering
      \includegraphics{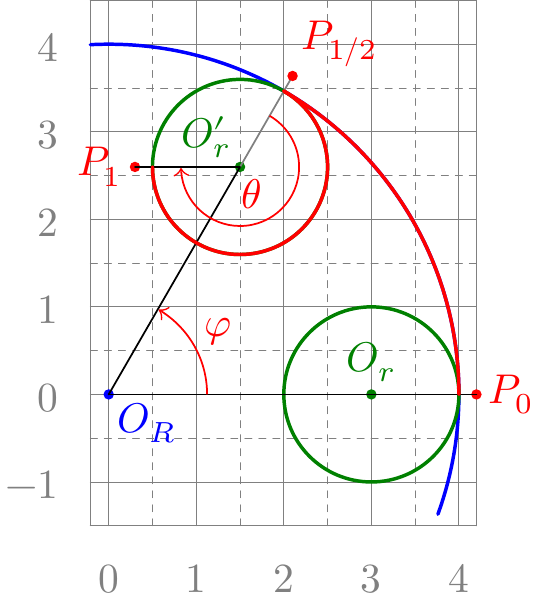}
      \caption{Гипотрохоида $d > r$}
      \label{fig:06}
    \end{minipage}
    \begin{minipage}[b]{0.32\textwidth}
      \centering
      \includegraphics{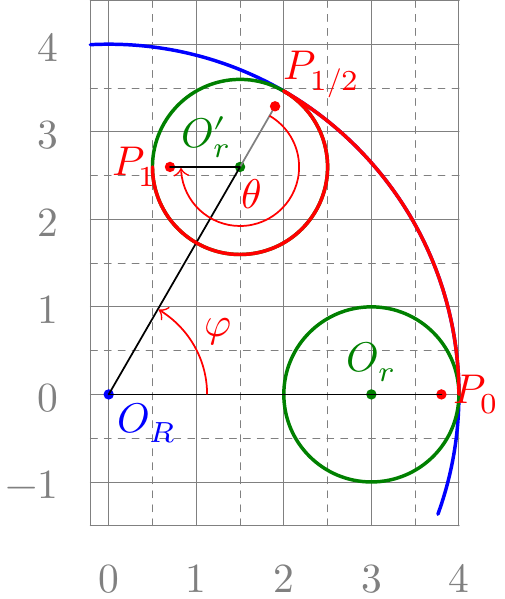}
      \caption{Гипотрохоида $d < r$}
      \label{fig:07}
    \end{minipage}
  \end{figure}

  Из определения гипотрохоиды и эпитрохоиды следует, что над точкой $P_0$  совершается движение $T(\varphi)$, состоящее из композиции двух поворотов~(рис.~\ref{fig:05}--\ref{fig:07}).
  \begin{enumerate}
    \item $T_1(\varphi)$ --- поворот вокруг точки $O_R$ на угол $\varphi$, при котором точка $O_r$ переходит в $O^{\prime}_{r}$, а точка $P_0$ в точку $P_{1/2}$.
    \item $T_2(\theta(\varphi))$ --- поворот вокруг точки $O^{\prime}_{r}$ на угол $\theta$, при котором $P_{1/2}$ переходит в $P_{1}$.
  \end{enumerate}

  \begin{figure}[h!]
    \begin{minipage}[b]{0.32\textwidth}
      \centering
      \includegraphics{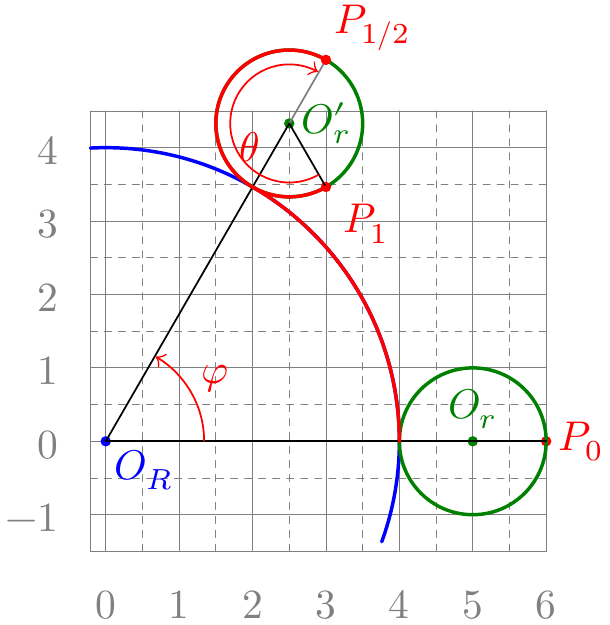}
      \caption{Эпициклоида $d = r$}
      \label{fig:08}
    \end{minipage}
    \begin{minipage}[b]{0.32\textwidth}
      \centering
      \includegraphics{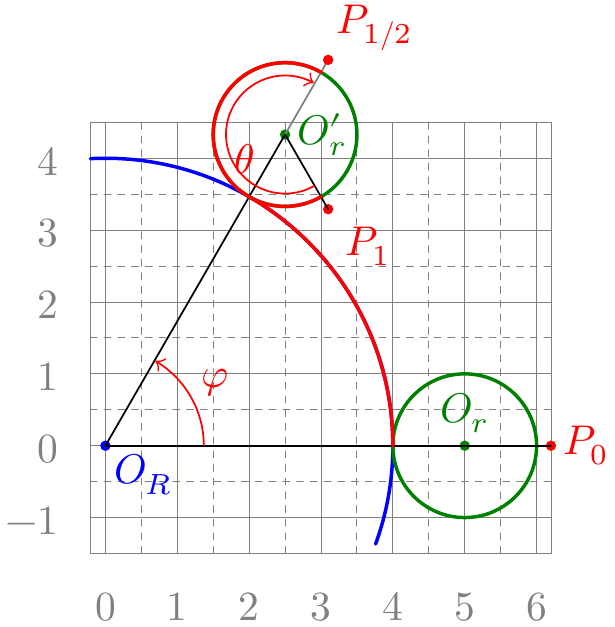}
      \caption{Эпитрохоида $d > r$}
      \label{fig:09}
    \end{minipage}
    \begin{minipage}[b]{0.32\textwidth}
      \centering
      \includegraphics{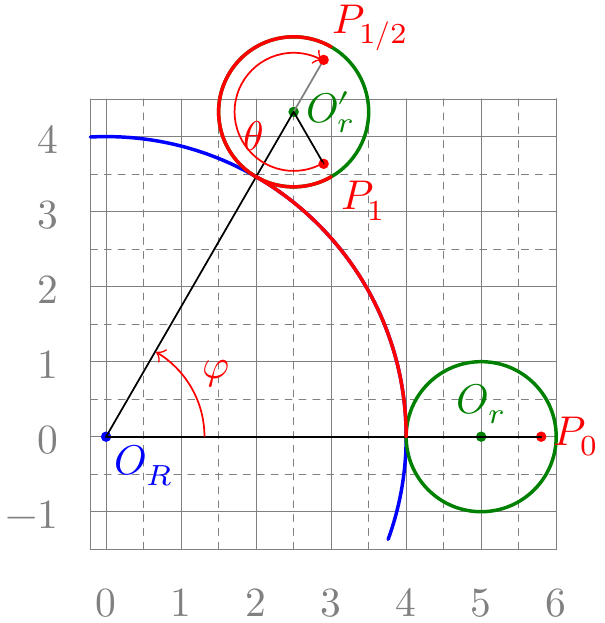}
      \caption{Эпитрохоида $d < r$}
      \label{fig:10}
    \end{minipage}
  \end{figure}

  Угол поворота $\theta$ связан с углом $\varphi$. Малая окружность должна прокатиться расстояние, равное длине дуги $PP_{1/2}$, что означает равенство длин дуг $PP_{1/2}$ и $P_{1/2}P_{1}$.
\begin{equation*}
  |PP_{1/2}| = R\varphi = |P_{1/2}P_{1}| = \theta r \Rightarrow \theta = \dfrac{R\varphi}{r} = \dfrac{\varphi}{k},\; k = r/R.
\end{equation*}

  Таким образом, для построения кривой достаточно задать параметры $R$, $r$ и $d$, начальные положения окружностей и точки $P_0$. Обычно полагают, что центр $O_R$ совпадает с началом координат, а центр $O_r$ лежит на оси $Ox$. Тогда координаты центра $O_r$ вычисляются как:
  \begin{equation}\label{eq:01}
    \vb{OO}_r = \vb{OO}_R + (R + s \cdot r, 0)^T,\quad
    s = 
    \left\{
      \begin{aligned}
        +1 & \text{, если эпитрохоида},\\
        -1 & \text{, если гипоциклоида}.\\
      \end{aligned}
    \right.
  \end{equation}
  А координаты точки кривой $P_0$:
\begin{equation*}
  \vb{OP}_{0} = \vb{OO}_r + (d, 0)^T.
\end{equation*}

  Теперь, чтобы найти любую току кривой, достаточно подействовать на $P_0$ преобразованием $T(\varphi) = T_2(\varphi/k) \circ T_1(\varphi)$ задав необходимое значение $\varphi$. Если необходимо построить множество точек, то взяв достаточно малый шаг $\varphi$, можно последовательно действовать на точку $P_0$ преобразованиями $T(i\varphi),i=1,2,\ldots,n$:
\begin{equation*}
  P_0 \stackrel{T( \varphi)}{\longrightarrow} P_1,\;
  P_0 \stackrel{T(2\varphi)}{\longrightarrow} P_2,\;
  P_0 \stackrel{T(3\varphi)}{\longrightarrow} P_3,\;
  P_0 \stackrel{T(4\varphi)}{\longrightarrow} P_4,\;,\ldots,
  P_0 \stackrel{T(n\varphi)}{\longrightarrow} P_n,\;.
\end{equation*}

  \section{Реализация на Asymptote}

Приведём далее исходный код программы на языке Asymptote и прокомментируем его ключевые моменты.

\newcounter{codlabel}
\newcommand{\cod}[1]{\refstepcounter{codlabel}(\thecodlabel)\label{#1}}

\begin{minted}{asymptote}
include "config.asy";

import animation;
import graph;

unitsize(1cm);
size(10cm, 10cm);

string ssign(int d) {
  return d > 0 ? "+" : "-";
}

transform T1(real phi, pair O_R) { // \cod{m1}
  return rotate(angle=phi, z=O_R);
}

transform T2(real phi, int sign, real k, pair O_r) { // \cod{m2}
  return rotate(angle=sign*phi/k, z=O_r);
}

pen BigCircle = blue;
pen littleCircle = deepgreen;
pen curve = red;

int sign = -1; // \cod{m3}
real R = 4.0;
real r = 2.0;
real d = 2.0;
real k = r/R;
int N = 100; // \cod{m4}
pair O_R = (0, 0);
int turns = 1; // \cod{m5}
usersetting(); // \cod{m6}

pair O_r = O_R + polar(R + sign * r, 0); // \cod{m7}
pair P = O_r + polar(d, 0);  // \cod{m8}
pair Q = O_r - sign * polar(r, 0);  // \cod{m9}

guide xcycloid;
transform T;

animation A; // \cod{m10}
A.global = true; 

draw(circle(c=O_R, r=R), p=BigCircle); // \cod{m11}
dot(O_R, p=BigCircle);


for(real phi: uniform(0, 360turns, N)) {
  save(); // \cod{m12}
  
  T = T1(phi, O_R) * T2(phi, sign, k, O_r); // \cod{m13}
  
  xcycloid = xcycloid -- T*P; // \cod{m14}

  draw(xcycloid, p=1bp+curve); // \cod{m15}
  dot(T*P, L=Label("P", align=NW)); // \cod{m16}
  draw(O_R -- T*O_r, L=Label("R"+ssign(sign)+"r")); // \cod{m17}
  draw(T*O_r -- T*P, L=Label("d"));
  draw(circle(c=T*O_r, r=r), p=littleCircle); // \cod{m18}
  dot(T*O_r, p=littleCircle);
  dot(T1(phi, O_R)*Q, p=littleCircle); // \cod{m19}

  include "axes.asy"; // \cod{m20}

  A.add(); // \cod{m21}
  restore(); // \cod{m22}
}

A.movie(); // \cod{m23}
currentpicture.erase();

\end{minted}

  Данная программа создает многостраничный PDF-файл, каждая страница которого является будущим кадром видео. Основную работу по вычислению точек кривой выполняют функции \verb|T1|~\eqref{m1} и \verb|T2|~\eqref{m2}. Эти функции определены для удобства, чтобы код максимально отражал приведенные выше формулы. Всю работу выполняет встроенная функция \verb|rotate|, которая позволяет определить вращение вокруг произвольной точки (аргумент \verb|z|) на произвольное значение угла в градусах (аргумент \verb|angle|).

  Далее задаем набор переменных-параметров~\eqref{m3}. Переменная \verb|sign| --- это $s$ из формулы~\eqref{eq:01}, а остальные соответствуют своим математическим обозначениям. Переменная \verb|N|~\eqref{m4} задает количество вычисляемых точек и, как следствие, кадров в будущем видеоролике. Переменная \verb|turns|~\eqref{m5} задает количество полных поворотов вокруг центра $O_R$. Вызов встроенной функции \verb|usersetting|~\eqref{m6} переопределить значение любой заданной выше переменной через аргумент командной строки \verb|-u|.
  
  Далее, на основе вышеопределенных параметров, вычисляются координаты начального положения центра движущейся окружности $O_r$~\eqref{m7}, точки кривой $P$~\eqref{m8} и точки касания $Q$ движущейся окружности с неподвижной~\eqref{m9}.
  
  Далее создается объект \verb|A|~\eqref{m10}, в который будут записываться кадры анимации (объекты типа \verb|picture| или \verb|frame|). У \verb|A| есть несколько полей, в частности поле \verb|global| типа \verb|bool| позволяет включать и отключать сохранение создаваемых изображений в виде массива в оперативной памяти и записи их в виде файлов на диск лишь после того, как они все будут построены.
  
  Вычисление точек кривой происходит в цикле, но перед этим рисуется неподвижная окружность~\eqref{m11} и ее центр. Затем вначале каждой итерации цикла сохраняются все текущие неподвижные элементы картинки (объект \verb|picture|)~\eqref{m12}, строятся все подвижные элементы, получившаяся картинка добавляется в структуру \verb|A|~\eqref{m21} и состояние изображения сбрасывается~\eqref{m22} в то, которое было на момент~\eqref{m12}. Процесс продолжается, пока не будут нарисованы и сохранены в \verb|A| все кадры.
  
  По мере продвижения цикла угл $\varphi$ изменяется от $0$ до $2\pi n$ (в градусах). На каждом шаге вычисляется преобразование поворота $T(\varphi)$~\eqref{m14}, применяется к начальной точке $P$ и добавляется к пути (\verb|guide|) \verb|xcycloid|~\eqref{m14}. С каждой итерацией цикла в путь \verb|xcycloid| добавляются новые точки и кривая растет.
  
  Далее следуют следующие команды рисования:
  \begin{itemize}
    \item уже вычисленной части кривой~\eqref{m15};
    \item нового положения точки $P$~\eqref{m16};
    \item отрезка длины $R + s\cdot r$~\eqref{m17}, соединяющего центр $O_R$ с новым положением центра $O_r$, а также отрезка длины $d$, соединяющего новый центр $O_r$ с точкой $P$ кривой;
    \item непосредственно саму движущуюся окружность в ее новом положении~\eqref{m18} и ее центр;
    \item точку касания $Q$~\eqref{m19};
    \item координатную сетку, настройки которой вынесены в отдельный файл~\eqref{m20}.
  \end{itemize}
  
  Наконец, после отработки цикла, все созданные кадры записываются в PDF файл. Для этого Asymptote последовательно создает отдельные PDF файлы для каждого кадра, затем добавляет на них текст, обработанный \LaTeX{} (в нашем случае Lua\LaTeX{}). Основное время работы программы занимает именно эта процедура, сами вычисления по сравнению с этим времени практически не занимают.
  
  Отметим также особенность синтаксиса Asymptote, которая позволяет опускать оператор \verb|*| при умножении числовых литеральных констант и переменных, например \verb|360turn|~\eqref{m12}.

  \section{Создание видеоролика}
  \subsection{Запуск Asymptote}

  Для запуска рассмотренной выше программы выполним следующую команду.
\begin{minted}[breaklines=true,linenos=false]{bash}
  asy -noV -nobatchView -f pdf -globalwrite -u 'R=3;r=1;d=1;N=100' xcycloid.asy -o video/xcycloid.pdf
\end{minted}

  Файл с исходным кодом \verb|xcycloid.asy| запускается на выполнение и в результате будет создан файл \verb|xcycloid.pdf|. Рассмотрим используемые опции.
  \begin{itemize}
    \item Опции \verb|-noV| \verb|-nobatchView| предотвращают автоматическое открытие вновь созданного изображения. Опция \verb|-noV| отключает эту функцию при выполнении из командной строки, а \verb|-nobatchView| при выполнении скрипта (как в нашем случае).
    \item Опция \verb|-f pdf| указывает, что следует сразу создавать PDF файл, минуя стадию postscript файла.
    \item Опция \verb|-globalwrite| дает возможность сохранить файл \verb|xcycloid.pdf| в любую директорию (в нашем случае \verb|video|), а не только в ту, где находится исходный файл \verb|xcycloid.asy|.
    \item Опция \verb|-u| позволяет взаимодействовать с функцией \verb|usersetting()| и передать внутрь программы значения переменных. Так мы передаем значения \verb|R=3|, \verb|r=1|, \verb|d=1| и \verb|N=100|. Такая возможность позволяет использовать один файл с исходным кодом для построения множества изображений, гибко настраивая любые параметры. Обратите внимание, что данный параметр принимает именно текстовую строку, которую затем обрабатывает функция \verb|usersetting()| поэтому передаваемые параметры необходимо брать в кавычки.
  \end{itemize}

  \subsection{Конвертация в PNG с помощью GhostScript}

  Для преобразования полученного многостраничного файла в видео формат, необходимо сконвертировать его страницы в растровые изображения. Для этого мы предлагаем использовать программу GhostScript~\cite{ghostscript}. Она доступна как для Windows, так и Unix систем (GNU/Linux, macOS). Также она поставляется с дистрибутивом \TeX{}Live~\cite{texlive}, как и Asymptote/

  Для конвертации PDF файла следует выполнить команду
\begin{minted}[breaklines=true,linenos=false]{bash}
  gs -sDEVICE=png16m -r600 -o video/xcycloid-%04d.png video/xcycloid.pdf
\end{minted}

  В случае, использования GhostScript из дистрибутива \TeX{}Live, следует вызывать \verb|gs| посредством скрипта \verb|rungs|, который находится
  \begin{itemize}
    \item в директории \verb|texlive\2023\bin\win32| в случае ОС Windows,
    \item в директории \verb|texlive/2023/bin/x86_64-linux| в случае GNU/Linux.
  \end{itemize}
  Каталог \verb|2023| соответствует версии дистрибутива \TeX{}Live и может отличаться.
  
  Параметр \verb|-sDEVICE=png16m| устанавливает формат изображения (полноценный PNG), параметр \verb|-r600| устанавливает плотность пикселей (dpi). Можно контролировать формат названия выходных PNG файлов с помощью символов форматирования, сходных с теми, которые используются в функции \verb|printf| языка Си. В нашем случае мы задали формат \verb|xcycloid-%04d.png|. Файлы будут пронумерованы начиная с \verb|0001| c 4 ведущими нулями.

  \subsection{Создание видео с помощью FFmpeg}

  Процесс склейки полученных растровых изображений в один видеоролик осуществим с помощью FFmpeg~\cite{ffmpeg:site}. Данная программа представляет собой утилиту командной строки и обладает обширным функционалом и, как следствие, огромным количеством опций и настроек. Приведем пример создания видео ролика из сгенерированных на предыдущем шаге PNG изображения и дадим пояснение к используемым параметрам.
\begin{minted}[breaklines=true,linenos=false]{bash}
  ffmpeg -r 30 -f image2 -start_number 1 -i video/xcycloid-%04d.png -c:v libx264 -vf "pad=ceil(iw/2)*2:ceil(ih/2)*2" video/xcycloid.mp4
\end{minted}

  \begin{itemize}
    \item Параметр \verb|-r| устанавливает частоту кадров.
    \item Параметр \verb|-f| устанавливает формат входного файла.
    \item Так как на вход подается множество файлов, следует указать формат их имен. Используются те же обозначения, что и в случае \verb|gs|. Параметр  \verb|-start_number| устанавливает начальный номер.
    \item Параметр \verb|-c:v| позволяет указать используемый видео энкодер. В нашем случае \verb|libx264|, но поддерживаются множество других форматов.
    \item Важный параметр \verb|-vf| задает фильтр, который накладывается на обрабатываемый кадр. В нашем случае мы округляем ширину и высоту кадра в пикселях до четного числа. После конвертации в PNG ширина и высота изображения могут оказаться нечетными, что недопустимо для подавляющего большинства энкодеров. Указанный фильтр позволяет избежать этой ошибки и перемасштабировать кадр силами ffmpeg.
  \end{itemize}
  На выходе мы получим видео, запакованное в контейнер \verb|mp4|. Выбранный нами формат x264 широко распространен и может быть воспроизведен любым современным браузером, не говоря уже о программах видеоплеерах.
\section{Заключение}
\label{sec:conclusion}

Мы подробно разобрали способ создания анимации векторной графики на плоскости с помощью языка Asymptote. Этот аспект данного языка в официальном руководстве освещен слабо и на наш взгляд данная статья в какой-то мере восполняет этот пробел. Хотя в результате получается видео-ролик, содержащий растровые изображения, но благодаря векторному исходнику (PDF) можно увеличивать разрешение видео практически безгранично.

Также следует заметить универсальность такого способа создания анимации, так как для создания набора изображений-кадров можно использовать практически любое средство визуализации данных. Всю работу по созданию видео файла делает FFmpeg.

\def\acknowledgmentsname{Благодарности}

\begin{acknowledgments}

Публикация выполнена при поддержке Программы стратегического
академического лидерства РУДН.

\end{acknowledgments}

  \ifthenelse{\boolean{xetex} \OR \boolean{luatex}}
  {\bibliographystyle{ugost2008l}}
  {\bibliographystyle{gost2008l}}

\bibliography{bib/asymptote-animation/cite}